\documentclass{article}
\usepackage{url}
\def\babar{\mbox{\slshape B\kern-0.1em{\smaller A}\kern-0.1em
    B\kern-0.1em{\smaller A\kern-0.2em R}}}
\newcommand{\cL}{{\cal L} }
\renewcommand{\aleph}{\mbox{{\em a}}}
\newcommand{\STi}{Slavnov-Taylor identity }
\newcommand{\WTi}{Ward-Takahashi identity }
\newcommand{\STis}{Slavnov-Taylor identities }
\newcommand{\WTis}{Ward-Takahashi identities }
\newcommand{\I}{\mathrm{i}}

\newcommand{\nn}{\nonumber}
\newcommand{\bary}{\begin{array}}
\newcommand{\eary}{\end{array}}

\newcommand{\gv}{\mbox{GeV}}

\newcommand{\ha}{\frac{1}{2}}

\newcommand{\semis}{\;;\;\;}
\newcommand{\cs}{\;,\;\;}
\newcommand{\epo}{\;. }

\newcommand{\be}{\begin{equation}}
\newcommand{\ee}{\end{equation}}
\newcommand{\ba}{\begin{eqnarray}}
\newcommand{\ea}{\end{eqnarray}}

\def\dah{\Delta\alpha^{(5)}_{\rm had}}

\def\dah0{\Delta\alpha^{(5)}_{\rm had}(-s_0)}

\usepackage{graphicx}  

\title{Comment on $H \to \gamma \gamma$ and the role of the decoupling theorem and
the equivalence theorem}
\author{F. Jegerlehner\\
Humboldt-Universit\"at zu Berlin, Institut f\"ur Physik,\\
       Newtonstrasse 15, D-12489 Berlin, Germany
}

\DeclareMathSymbol{\varPhi}{\mathalpha}{operators}{"08}
\DeclareMathSymbol{\varOmega}{\mathalpha}{operators}{"0A}

\begin{document}
\renewcommand{\thefootnote}{\fnsymbol{footnote}}
\setlength{\baselineskip}{0.52cm}
\thispagestyle{empty}
%
%
%

\maketitle


\begin{abstract}
I am commenting on the recent paper~\cite{Gastmans:2011ks} about a new
calculation of the $H\to\gamma\gamma$ rate mediated by the $W$ boson
loop, and the lack of decoupling of the heavy states which mediate the
decay. We remind the reader that the heavy Higgs limit is dominated by
the contribution from the longitudinal $W$ bosons, which in the limit
$m_H \gg M_W$ are represented by the charged Higgs ghosts according to the
equivalence theorem. The corresponding contribution is missing
in~\cite{Gastmans:2011ks}.
\end{abstract}

\section{Introduction}
In a recent paper~\cite{Gastmans:2011ks} previous one-loop
calculations of the decay $H\to \gamma \gamma$ via the $W$ boson loop
in the electroweak Standard Model (SM) have been criticized to be
incorrect. In~\cite{Gastmans:2011ks} it is argued that in the limit
$m_H\gg M_W$ the Higgs decay amplitude $A^{(W)}_{H\gamma\gamma}$
should stay bounded, while it is actually $\propto m_H^2$, a behavior
which is claimed to be violating the decoupling theorem. In this short
note we defend previous calculations and explain why previous results
are correct. The main point is that the decoupling theorem~\cite{AC75}
is a statement about the limit $M_W \gg m_H$ while the limit $m_H \gg
M_W$ is ruled by the equivalence theorem~\cite{ET74}. It is well known
that heavy Higgs and heavy top physics has little to do with the gauge
sector (since the heavy Higgs, heavy top effective theory is there for
vanishing gauge couplings $g,g'=0$), but is determined entirely by the
symmetry breaking sector, the Higgs and the Yukawa
sector~\cite{Barbi}, 
\ba
\cL_{\rm eff}&=& 
\partial_\mu\varPhi^+ \partial^\mu \varPhi
+\bar{t}\,\gamma^\mu \partial_\mu\,t+\bar{b}\,\gamma^\mu \partial_\mu\,b
+\mu^2\,\varPhi^+\varPhi+\lambda \left(\varPhi^+\varPhi \right)^2
\nn \\ && -y_t\left(\bar{Q}_L\varPhi^c t_R+\mathrm{h.c.}\right)
\ea
where covariant derivatives appear replaced by normal
derivatives and the local $SU(2)_L\otimes U(1)_Y$ symmetry 
is replaced by a global
$SU(2)_L$ symmetry with corresponding \WTis$\!$. $\varPhi$ is the
$Y=1$ Higgs doublet field
\ba
\varPhi(x)=\left(\begin{tabular}{c} $\varphi^+$ \\ $\varphi^0$  \end{tabular} \right)\semis \varphi^0=\frac{H+v-\I\,\varphi}{\sqrt{2}}\epo
\ea
By $\varPhi^c=\I\,\tau_2\varPhi^*$ we denote the $Y$ charge conjugate
$Y=-1$ Higgs doublet.  $Q_L$ is the left-handed $(t,b)$ doublet. The
$b$ as a light field can be taken to be non-interacting. In our case
of $H\to 2\gamma$ photons couple as usual to the charged particles in the
heavy sector (charged Higgses and the top and bottom
quarks). Furthermore, the limit $g,g'\to 0$ should be taken at fixed
low energy constraint on $\sin^2 \Theta_W=
1-\frac{M_W^2}{M_Z^2}=\frac{g'^2}{g^2+g'^2}$. The adequate
renormalization scheme takes $G_\mu$ and $\sin^2 \Theta_W$ as input
parameters together with $\alpha$ for the QED part.

We first look at the decoupling limit: the \textbf{decoupling theorem} states
that heavy virtual particles of mass $M$ decouple like $O(E/M)$ as
$M\to \infty$ where $E$ is the fixed energy or light-mass scale of the
``light'' particle sector. In fact the Appelquist-Carrazone
decoupling--theorem~\cite{AC75} holds in theories like QED and QCD
only, where masses and couplings are independent and when some of the
masses get large at fixed couplings.  In the SM where masses are
generated by the Higgs mechanism the decoupling theorem does not hold
in general because of the well known mass coupling relations
\ba
M_W=\frac{g\,v}{2}&\cs &M_Z=\frac{g\,v}{2\cos \Theta_W} \nn \\ 
m_f=\frac{y_f\, v}{\sqrt{2}} &\cs & m_H=\sqrt{2\lambda}\,v \epo
\ea
In the SM masses can only get large either in the strong coupling
regime, or by taking the Higgs vacuum expectation value $v
\to \infty$, which then would violate the important low energy constraint
\ba
v=\left(\sqrt{2}\,G_\mu\right)^{-1/2}=246.2186(16)~\gv \cs
\ea  
and in addition would rescale the spectrum uniformly to large masses.  In
the SM a particle cannot be removed from the theory by just taking its
mass to infinity. At fixed $v$ it requires to take the strong coupling
limit in which perturbative arguments fail to apply.

One drawback, relevant for the $H \to \gamma \gamma$, is that the
$HWW$ coupling is $2M_W^2/v$, and similarly, the $H\bar{\psi}_f\psi_f$
fermion couplings are $m_f/v$.  As a consequence, masses appear as
factors in the numerators of Feynman amplitudes in addition to the
masses in propagators [mass terms] which show up in the denominators.
The mass factors in the numerators obviously lead to non decoupling
effects for large masses.

In discussing various mass limits in the SM one should keep in mind
that relations like the custodial symmetry constraint
$\rho=\frac{M_Z^2}{M_W^2\,\cos^2 \Theta_W}=1$ or equivalently $\cos^2
\Theta_W=M_W^2/M_Z^2$ must be respected. In the gauge field sector we
have three basic parameters, the gauge couplings $g$, $g'$ and the
Higgs vacuum expectation value $v$ which usually (LEP parametrization)
are mapped to $\alpha$, $G_\mu$ and $M_Z$ as the most precisely known
input parameters. Then the $W$ mass is given by
\ba
M_W^2/M_Z^2=\frac{1}{2}\,\left(1+\sqrt{1-\frac{4\,A_0^2}{M_Z^2}\,\frac{1}{1-\Delta
r}}\right) =\cos^2\Theta_W
\label{MWMZrelation}
\ea
considered to be kept fixed.
Here $A_0=\left(\frac{\pi \alpha}{\sqrt{2}\,G_\mu}\right)^{1/2
}=37.2802(3)~\gv$ and $\Delta r$ represents known radiative
corrections. In any case we must respect $M_W < M_Z$, or more
precisely the relation (\ref{MWMZrelation}). We just wanted to point
out that non-decoupling effects in the SM are experimentally well
established, and one cannot conclude a SM calculation to be
incorrect because of lack of decoupling. Of course the lack of
decoupling naively looks unnatural, but in fact constraints from the
$\rho$--parameter on heavy states are quite intriguing and extensions
of the SM in most cases end up in a fine tuning
problem~\cite{Czakon:1999ha}, because decoupling of new heavy states,
in theories where masses are generated by spontaneous symmetry
breaking, is more the exception than the rule. Therefore, experimental
low energy constraints actually are much more severe than often
anticipated.

The \textbf{equivalence theorem} has to do with the massive gauge bosons
in the limits where gauge boson masses are expected to become
irrelevant.  This concerns high energies $E \gg M_W,M_Z$ as well as
the effective regime when gauge boson masses are small relative to
other masses, like $m_t,m_H \gg M_W,M_Z$. Specifically for the heavy
top effects this has been worked out and discussed in~\cite{Barbi}
and~\cite{Fleischer:1994cb}.
 
The root of the equivalence theorem is the following: in the unbroken
phase of the SM the gauge bosons are massless and have two transverse
degrees of freedom.  When the SM undergoes spontaneous symmetry
breaking (Higgs mechanism) the gauge bosons become massive and get a
third physical degree of freedom, the longitudinal one, by ``eating up
the Higgs ghosts'', which become unphysical. The number of physical
degrees of freedom remains conserved.  The broken symmetry is
recovered as an asymptotic symmetry when energies get large relative
to the gauge boson masses. Equivalently, in the limit $M_W,\,M_Z\ll E$
the gauge bosons become transversal again, however, the longitudinal
modes transmute back into physical scalar Higgs degrees of
freedom. Three of the four scalar Higgses had turned into unphysical
Higgs ghosts during symmetry breaking. At high energies the symmetry
is restored. 

Formally, one may use 't Hooft's gauge fixing conditions (linear
covariant gauge)
\ba
\begin{array}{l c l c}
W_{\mu}^{\pm} & : & C^{\pm} = - \partial _{\mu}W^{\mu \pm} \pm  i \xi _W M_W 
\varphi ^{\pm} & = 0 \\
Z_{\mu} & : & C_Z = - \partial _{\mu} Z ^{\mu} - \xi _Z M_Z \varphi &  = 0 \\
A_{\mu} & : & C_A = - \partial _{\mu} A ^{\mu}  & = 0 \\
\end{array}
\label{gaugefix}
\ea
which imply a correspondence like $\partial _{\mu}W^{\mu \pm} \propto
\varphi ^{\pm}$.  While the transversal modes couple with gauge
coupling $g\,HW^+W^-$ the ghosts couple with the Higgs self coupling
$\lambda\,H\varphi^+\varphi^-$ and when $m_H \gg M_W$, meaning $\lambda
\gg g$, the longitudinal modes dominate. This is the limit referred to
in Ref.~\cite{Gastmans:2011ks} as decoupling limit, mistakenly the
dominating longitudinal mode has been lost in the calculation.

The relevant \WTis derive from the standard model \STis as follows. We
use the notation of Ref.~\cite{TASI}. In the 't Hooft gauge we denote
by $\xi$ the gauge parameter, $\aleph$, $\zeta$ and $\eta^\pm$ are the
photon-associated, the neutral and charged Faddeev--Popov ghost
fields, respectively. The $W$ boson propagator satisfies
\ba
&&  < T \partial _{\mu} W ^{\mu+} (x) W^-_{\nu} (y) >
+ \xi M_W < T \varphi^+ (x) W^-_{\nu} (y) >  \nn \\
&= & - \xi < T \bar{\eta}^+ (x) \partial _{\nu} \eta^- (y) > 
+ \I\,\xi\,[e\,< T \bar{\eta}^+ (x)\,(W^-_\nu\aleph-A_\nu\,\eta^-)(y)>
\nn \\&&-g\; \cos \Theta_W\,< T \bar{\eta}^+ (x) ( W^-_{\nu} \zeta - Z_{\nu} \eta^-) (y) >]
\nn
\ea
and
\[
\begin{array}{l l}
& < T \partial _{\mu} W ^{\mu+} (x) \partial _{\nu} W^{\nu-} (y) >
+\; \xi M_W < T
\partial _{\mu} W^{\mu+} (x) \varphi^- (y) > \\
& +\; \xi M_W < T \varphi^+ (x) \partial _{\nu} W ^{\nu-} (y) >
+\; \xi ^2 M_W^2 < T \varphi^+ (x) \varphi^- (y) > = - \I\, \xi \delta (x - y)
\end{array}
\]
for the longitudinal parts of the gauge field propagators.
Using the usual tensor decomposition for the self-energy functions
(inverse propagators) in Fourier space
\[
\begin{array}{ccccc}
< T W^{\mu+} (x) W^{\nu-} (y) >       &\to &
\I\,     \left(g^{\mu \nu} A_1 + q^\mu q^\nu A_2 \right) &\equiv &
-\I\, \left(g^{\mu \nu} \Pi_W(q^2)+\cdots \right) \\
< T W^{\mu+} (x) \varphi^- (y) >     &\to &   M_W p^\mu B_1   &       & \\
< T \varphi^+ (x) \varphi^- (y) >   &\to & \I\,  C_1            &\equiv &
 \I\, \Pi_\varphi (q^2) \\
< T \bar{\zeta}^+ (x) \zeta^- (y) > + \cdots &\to & -\I\, M_W^2  D_1       &       &
\end{array}
\]
the above identities read:
\[
\begin{array}{r}
\xi \:\left(A_1+q^2 A_2+ \;\; B_1 \right)+ D_1 =0\;\; \\
q^2 \:\left(A_1+q^2 A_2+ 2 B_1 \right)+ C_1 =0\;.
\end{array}
\]
By $D_1$ we have denoted the full Faddeev--Popov ghost contribution
which includes the three terms on the r.h.s. of the first of the above \STis.

In the limit $g \to 0$ the Faddeev-Popov ghosts do not contribute and
therefore $D_1 \simeq 0$ such that $A_1+q^2 A_2+ B_1 \simeq 0$. Since the
self--energy amplitude $A_2$ does not exhibit a pole at $q^2=0$ we have
$q^2 A_2 \to 0$. Thus we obtain the relevant \WTi
$A_1 \simeq \frac{C_1}{q^2}$ which we may write in the form
\be
\frac{\Pi_W(q^2)}{M_W^2} \simeq -\frac{\Pi_{\varphi^\pm} (q^2)}{q^2}
=-\Pi'_{\varphi^\pm} (q^2)
\ee
and which expresses the physical transversal part of the $W$
self--energy in terms of the self--energy of the charged scalar Higgs
ghosts. For the $HW^+W^-$ vertex the \STis look the same
as for the $W^+W^-$ propagator with a Higgs field in addition under
the time-ordering prescription: like
\ba
\langle T \partial_\mu W^{\mu+} W^{\nu-} H\rangle-\I\,\xi\,M_W\,\langle
T \varphi^+ W^{\nu-} H\rangle=-\xi\langle T \bar{\eta}^+\partial_\mu
\eta^- H \rangle + \cdots
\ea 
modulo the gauge variation of the Higgs field, which vanishes when the
Higgs is taken on-shell. Again in the limit of vanishing gauge couplings the r.h.s is
vanishing, meaning that the longitudinal component of the $W$ is
replaced by its charged Higgs ghost.

For illustration of a similar mechanism we remind about another
example of non-decoupling and the play of the equivalence theorem: the
heavy top contribution to the $W$ self--energy. In the limit $m_t \gg
M_W,M_Z$ the gauge coupling $g\,W^{\mu+}(\bar{t}\gamma_\mu\,\Pi_-\,b)$
turns into $y_t\,\varphi^+ (\bar{t}\,\Pi_-b)-y_b\,\varphi^+(\bar{t}\,\Pi_+b)$
($\Pi_\pm=(1\pm\gamma_5)/2$ the chiral projectors). Most prominent
example of an electroweak non-decoupling heavy top effect is the well
known low energy effective neutral to charged current coupling ratio
$\rho=G_{\rm NC}/G_{\rm CC}$. It gets renormalized as
$\rho=1+\Delta \rho$ where $\Delta \rho=
\frac{\Pi_Z(0)}{M_Z^2}-\frac{\Pi_W(0)}{M_W^2}\simeq
\Pi'_{\varphi^\pm}(0)-\Pi'_\varphi(0)=
\frac{\sqrt{2}G_\mu\,N_c}{16
\pi^2}\,|m_t^2-m_b^2|$ at one loop. This leading top effect is 
Veltman's ``flag pole''~\cite{Veltman:1976rt} and allowed to
``measure'' the top mass indirectly at LEP~\cite{:1993ac}, prior to
the direct top discovery at the Tevatron~\cite{topquark}. Similarly,
$B-\bar{B}$ oscillations discovered by Argus at
DESY~\cite{Albrecht:1987dr} were possible because the effect is
enhanced by a non-decoupling heavy top contribution (see
e.g.~\cite{Gaillard:1974hs,Buras:1984pq}).  Non-decoupling effects in
extensions of the SM have been discussed in
Ref.~\cite{Czakon:1999ha}. In the context of Higgs production and
decay non-decoupling phenomena of heavy fermions where investigated
long time ago in~\cite{Fleischer:1982us} (see
also~\cite{Fleischer:1982af,Jegerlehner:2005gf}). Last but not least,
the present indirect Higgs mass bound from LEP is due to a
non-decoupling effect. When we try to remove the virtual Higgs from
the SM by increasing its mass the SM would turn into a
non-renormalizable theory as we know. In fact $m_H^2$ effects in the
limit $m_H \gg M_W,M_Z$ at one loop are screened by the custodial
symmetry of the minimal Higgs system and only a logarithmic Higgs mass
dependence persists~\cite{Veltman:1994vm} in this case.

In summary: in the limit under
consideration physical S-matrix elements are dominated by the longitudinal
vector boson degrees of freedom and according to the {\em equivalence
theorem}, with $m_H\, [m_t]$ as a high energy scale, one
is allowed to replace (up to a phase and up to $O(M/m_H)\, [O(M/m_t)]$ corrections)
a longitudinally polarized vector boson by its corresponding
unphysical scalar.  An equivalent relationship is obtained in the
limit of vanishing gauge couplings, $g',\:g\to 0$, from the \WTis which
derives from the remaining global symmetry~\cite{Barbi}. 

\section{$H\to 2\gamma$ at one-loop}
For $H\to\gamma\gamma$ the correct SM results are well known.  The
$W$-loop amplitude is (see
e.g.~\cite{Ellis:1975ap,Shifman:1979eb,Fleischer:1980ub,Marciano:1987un})
\ba
A^{(W)}_{H\gamma\gamma}&=&C_W\, \left[3\,C_0(M_W,M_W,M_W;0,0,m_H^2)\,(2\,M_W^2-m_H^2)-3-\ha\,x_W\right]
\nn \\ &\stackrel{x_W \to 0}{\sim}& C_W\,\left[ 
\frac{7}{4}\,x_W+\frac{11}{120}\,x_W^2+\frac{19}{1680}\,x_W^3 +\cdots\right]
\nn \\ &\stackrel{z_W \to 0}{\sim}&3 C_W\,\biggl\{\biggl[
\fbox{$\frac{1}{6 z_W}$}+\left(\pi^2-\ln^2
z_W+2\right)/2  \nn \\ && \hspace*{6mm}
-\left(\pi^2-\ln^2 z_W+2\,\ln z_W\right)\,z_W
+\left(\ln z_W-2\right)\,z_W^2+\cdots\biggr] \nn \\ && \hspace*{6mm}  
+\I\,\pi\,\left[-\ln z_W +2\,\left(\ln z_W-1\right)\,z_W+z_W^2 +\cdots\right]
\biggr\}
\ea
with $C_W=\frac{\alpha}{2\pi}\,M_W^2$ and
$C_0(M,M,M;0,0,s)=\frac{2}{s}\,\left(\arctan
\frac{1}{\sqrt{4M^2/s-1}}\right)^2$ for $s \leq 4M^2$. Furthermore, we
denoted $x_W=m_H^2/M_W^2$ and $z_W=1/x_W$. For $s > 4M^2$ we have
$C_0=-\frac{1}{2s}\left(\ln
\frac{1-\sqrt{1-y}}{1+\sqrt{1-y}}+\I \pi\right)^2$ with $y=4M^2/s$. 
For fixed $HWW$ coupling $C_W$ is fixed and the amplitude exhibits
decoupling i.e. it is $O(x_W)$ as $x_W \to 0$. Taking into account the
growth of the coupling, however, the complete amplitude for $m_W \gg m_H$ tends
to a constant and lacks decoupling in the naive sense. The other limit
$z_W \to 0$, i.e. $m_H \gg M_W$, is exhibiting the singular term in
the box. This term is the one which has been questioned
in~\cite{Gastmans:2011ks}. The equivalence theorem requires this term
to be there without question. 

We have calculated these amplitudes in the 't Hooft
gauge with an arbitrary gauge parameter $\xi$ as well as in the unitary gauge
using dimensional regularization. The off-shell Slavnov-Taylor identities have been
checked as well to be satisfied.  In the 't Hooft gauge with free
gauge parameter $\xi$ there are 13 diagrams contributing. In the
unitary gauge there are 2\footnote{In a renormalizable gauge each $W$
is represented by a $W$ or $\varphi$ line which yields $2^3+2^2=12$
diagrams plus the Faddeev-Popov ghost loop.  }:
\begin{figure}[h]
\centering
\includegraphics[]{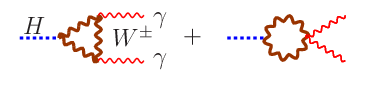}
\end{figure}

\noindent
Both calculations give identical results\footnote{In the unitary gauge
some technicalities with so called Lee-Yang terms~\cite{LeeYang} must
be taken into consideration also if dimensional regularization is
applied.}.

Using the equivalence theorem, we may
calculate the leading contribution for $m_H \gg M_W$ by calculating
the much simpler diagrams 
\begin{figure}[h]
\centering
\includegraphics[]{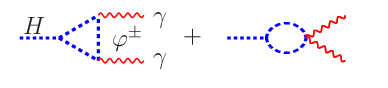}
\label{HggPloop}
\end{figure}

\noindent
exhibiting loops of the charged Higgs ghosts only.  The result in the
't Hooft gauge with arbitrary gauge parameter $\xi$ reads
\ba
A^{(\varphi)}_{H\gamma\gamma}&=&C_W\, \left[-\ha\,x_W+
C_0(M_\varphi,M_\varphi,M_\varphi;0,0,m_H^2)\,M_\varphi^2\,x_W\right]
\nn \\ &\stackrel{M_W \to 0}{\sim}&3 C_W\,\biggl\{
\fbox{$\frac{1}{6 z_W}$}+O(M_W^2/m_H^2)
\biggr\}\epo
\ea
In the full SM in the 't Hooft gauge $M^2_\varphi=\xi M^2_W$ is gauge
dependent, however, in the limit where the equivalence theorem applies
we have $M_\varphi/m_H \to 0$ where the second term vanishes. The
remaining physical (gauge invariant) leading term agrees
precisely with the leading term of the full SM calculation. The
subleading term is gauge dependent and hence unphysical. We conclude that
physics uniquely fixes that questioned leading term, and it also
implies that Slavnov-Taylor (ST) identities are obviously not
respected in the calculation of Ref.~\cite{Gastmans:2011ks}.

For comparison, the corresponding result for a heavy top loop is given
by
\ba
A^{(f)}_{H\gamma\gamma}&=&2\,Q_f^2\,C_f \left[ 
1+C_0(M_f,M_f,M_f;0,0,m_H^2)\,(\ha \,m_H^2-2\,M_f^2)\right]
\nn \\ &\stackrel{x_f \to 0}{\sim} &2\,Q_f^2\,C_f \left[ 
\frac{1}{6}\,x_f+\frac{7}{720}\,x_f^2+\frac{1}{1008}\,x_f^3+\cdots\right]
\nn \\ &\stackrel{z_f \to 0}{\sim} &2\,Q_f^2\,C_f \biggl\{\left[ 
1+\frac{1}{4}\left(\pi^2-\ln^2 z_f\right)
\right. \nn \\ && \left.\hspace*{6mm}
-\left(\pi^2-\ln^2 z_f+\ln z_f\right)\,z_f
 +\left(\frac{5}{2}\,\ln z_f-1\right)\,z_f^2+\cdots\right] 
 \nn \\ && \hspace*{6mm}
+\I\, \pi \,\left[ -\frac{1}{2}\,\ln z_f
+\left(2\,\ln z_f-1\right)\,z_f+\frac{5}{2}\,z_f^2+\cdots\right]
\biggr\}
\ea
with $C_f=\frac{\alpha}{2\pi}\,M_f^2$, $x_f=m_H^2/M_f^2$ and
$z_f=1/x_f$. 
\begin{figure}
\centering
\includegraphics[height=7.5cm]{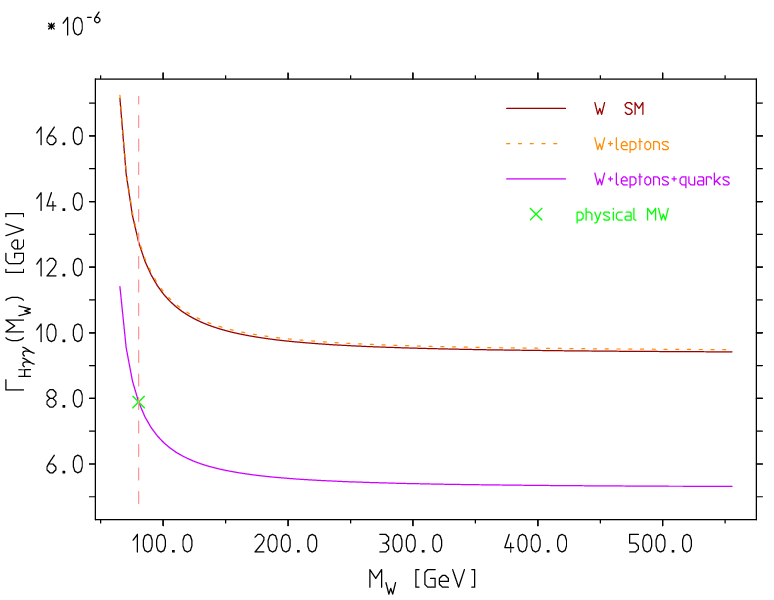}
\caption{The $H \to \gamma \gamma$ width as a function of the $W$ mass
at one loop order ($m_H=120~\gv$, $m_t=171.3~\gv$).}
\label{fig:Hgg} 
\end{figure}
\begin{figure}
\centering
\includegraphics[height=7.5cm]{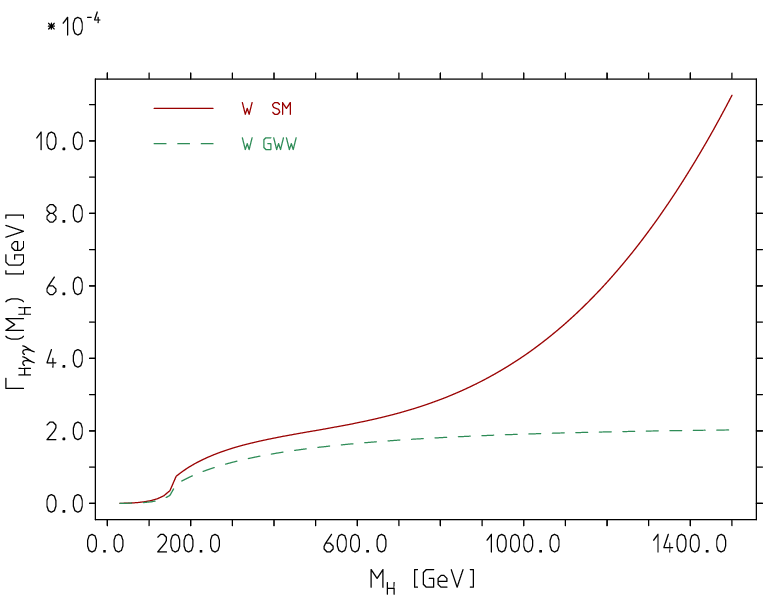}
\caption{The partial Higgs width $\Gamma(H \to \gamma \gamma)$ as a
function of the Higgs mass $m_H$. The full line is the standard SM
result, the dashed one the one calculated in Ref.~\cite{Gastmans:2011ks}.
}
\label{fig:Hggalt} 
\end{figure}

\begin{figure}
\centering
\includegraphics[height=7.5cm]{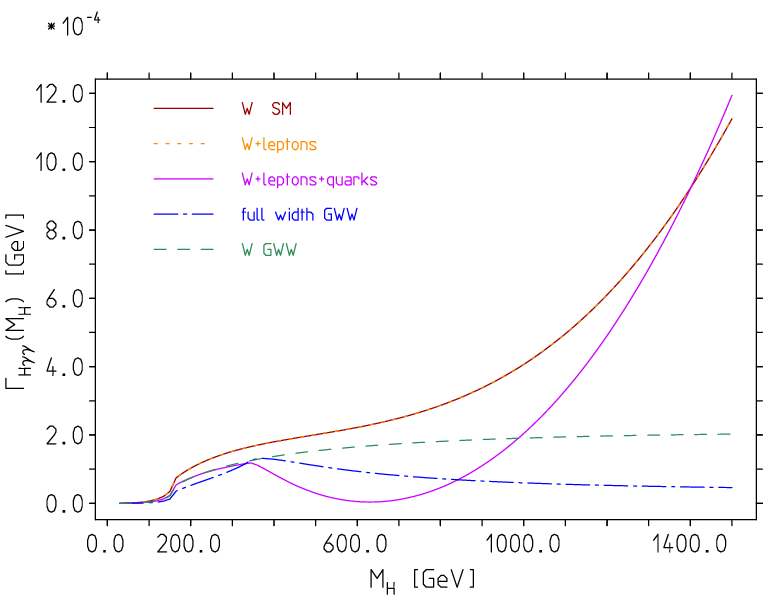}
\caption{Comparison of the $\Gamma(H\to\gamma\gamma)$ standard SM
predictions ($W$ loop only, + leptons, + quarks) with the alternative
prediction from Ref.~\cite{Gastmans:2011ks}}
\label{fig:compare} 
\end{figure}

The $H \to \gamma \gamma$ width is given by
\ba
\Gamma_{H\gamma\gamma}=\frac{\sqrt{2}\,G_\mu}{4\pi\,m_H}\,|{\cal A}_{H\gamma\gamma}|^2
\ea
where ${\cal
A}=A^{(W)}_{H\gamma\gamma}+\sum_f\,A^{(f)}_{H\gamma\gamma}$.
Figure~\ref{fig:Hgg} shows how the $H \to \gamma \gamma$ partial width
as a function of $M_W$ for $M_W \to \infty$ tends to a constant.
Figure~\ref{fig:Hggalt} compares the $W$ mediated Higgs width in the
heavy Higgs limit, with and without the proper leading term. In
Fig.~\ref{fig:compare} we finally compare the full SM prediction with
the $W$ mediated result, all in one-loop approximation. By GWW we
denoted the result from Ref.~~\cite{Gastmans:2011ks}.

The decay $H \to 2\gamma$ has been investigated in great details
including higher order effects. The two-loop QCD corrections to the
quark loops have been presented
in~\cite{Dawson:1992cy,Melnikov:1993tj,Djouadi:1995gt,Fleischer:2004vb}.
The limit $m_H >> M_W$ has been investigated at the two loop level
in~\cite{Korner:1995xd}. The complete two-loop corrections are also
known~\cite{Aglietti:2004nj,Degrassi:2005mc}. Not surprisingly, they
get large in the heavy Higgs range where strong coupling problems show
up.  For a comprehensive overview of the Standard Model Higgs profile
see~\cite{Djouadi:2005gi} and references therein.

Comments like the ones presented in this note have been presented in
Refs.~\cite{Shifman:2011ri,Huang:2011yf,Marciano:2011gm,SZC} in response
of~\cite{Gastmans:2011ks}. In our view, the result in question may be
interpreted as follows: in the regime $m_H \gg M_W$ only the
contribution from the massless transversal $W$'s have been taken into
account, while the contribution from the [in this limit] physical
charged massless Higgses has not been taken into account.

We conclude that the symmetry properties of the SM uniquely fix the
correct answer for $H \to 2\gamma$ rate. Dimensional regularization
(DREG) remains the most adequate technical tool to preserve the gauge
symmetry properties in calculations exhibiting ultraviolet divergent
integrals at some stage of a calculation. This even is so if no
overall renormalization is required like in the $H \to 2\gamma$
process. Whatever regularization prescription is utilized at the end
one has to make sure that the Slavnov-Taylor and \WTis are respected.
Limitations of dimensional regularization are well known in connection
with chiral structures: the anticommuting $\gamma_5$ problem and the
related Adler-Bell-Jackiw triangle anomaly (see
e.g.~\cite{Jegerlehner:2000dz} and references therein). Similarly, for
supersymmetric structures dimensional reduction (DRED) is a more
adequate~\cite{Stockinger:2005gx} regularization procedure. These
technicalities however do not play a role in the SM $H \to 2\gamma$
decay.
 
Because of the importance of the result in view of the Higgs
search at the LHC I find it appropriate to publicize these comments in
spite of the fact that essentially only known results are reviewed.

\section{Addendum and Update concerning Ref.~\cite{Christova:2014mea}}
A recent paper entitled ``Once more the $W$-loop contribution to the
Higgs decay into two photons'' finds a confirmation of the
calculations presented in Refs.~\cite{Gastmans:2011ks}, which
disagreed with earlier results obtained by utilizing the dimensional
regularization (DREG) approach. This result has been criticized by a
number of papers~\cite{Shifman:2011ri,Huang:2011yf,Marciano:2011gm},
including the original version of this note~\cite{J} form October
2011. More recently the problem has been addressed also in
Refs.~\cite{SZC,CCNS,DP}. The new calculation is based on applying the
Cutkosky rule together with a Dispersion Relations (DR) avoiding in
this way the need for a ultraviolet (UV) regularization. While the
imaginary part is is devoid of UV problems at leading order, the
application of dispersion relations in electroweak theory can be
tricky. This has been elaborated in~\cite{Kniehl:1991gu}, which was
rectifying results presented in Ref.~\cite{Kniehl:1988ie} for the
gauge boson self-energy functions. Specifically, the contributions to
the neutral to charged current Fermi type coupling, the so called
$\rho$-parameter, exhibits a leading correction $\Delta \rho (0)=
\frac{\Pi_{ZZ}(0)}{M_Z^2}-\frac{\Pi_{WW}(0)}{M_W^2}$, where both self-energy
functions are quadratically divergent, while the difference is
super-convergent, when represented as a dispersion
integral (DI). Nevertheless, the result in general need not be
correct, which at first looks really like a true surprise.
Indeed, the DI representation does not respect
\WTis and or \STis as given by (\ref{gaugefix}) automatically! If we
calculate the integrals before taking the difference, in the example
of $\Delta \rho$, we need a UV cut-off
and such cut-offs in case of non-Abelian gauge theories violate
\STis like (\ref{gaugefix}). Unlike in the case of photons,
where the physics is confined in the transversal part, for massive
gauge bosons there is a relation between the transversal and the
longitudinal part which must be respected and in an actual calculation
must be checked! Indeed, the weak point of the present paper is the
sentence ``Adding a constant term to the dispersion integral is of
course possible, as always, but does not seem to be justified by any
physical requirement. (We do not view the agreement with a calculation
using dimensional regularization as a physical requirement.)''. Well,
dimensional regularization is precisely designed to respect the
symmetries of the theory and the statement that this is not a physical
requirement is certainly not tenable. That's precisely the point:
choosing the constant to be zero is ambiguous and is just violating
the equivalence theorem, which relates the longitudinal $W$-mode to
the corresponding Higgs scalars ghost $\phi$. The Higgs scalars get
back their physical meaning in the symmetric phase of the SM, which
represents the high energy limit of the broken phase. It looks as if
the $W$'s in the calculation presented in~\cite{Christova:2014mea}
exhibit the transversal modes only. For large Higgs masses $m_H\gg M_W$
the rate must be dominated by the longitudinal modes which there are
represented by the scalar Higgs ghosts as given by Fig.~\ref{HggPloop} above. The
problem is not the electromagnetic
\WTi but the \STi for the divergence of the $W$-field. It is this
relation which fixes the otherwise arbitrary constant in the DR.
Likely, the flaw lies in choosing the unitary gauge where \STis of the
't Hooft gauge form (\ref{gaugefix}) get singular and the derivation
of the equivalence theorem gets less transparent. Although in the
unitary gauge we should get the correct answer of course, the
calculation is more cumbersome for what concerns respecting local
gauge invariance. I guess that the non-transversal part gives an
additional contribution. Also: in the unitary gauge in any case
possible UV singularities are worse than in a renormalizable
gauge. The \STis allow us to understand why the Higgs coupling 
$HW^+W^-$ which is proportional to $2M_W^2/v\sim g^2v/2 $ in the regime $m_H \gg
M_W$ must transmute into a coupling $H\varphi^+\varphi^-$ which is
proportional to $m_H^2/v\sim \lambda
v/3$. The Higgs ghosts which have sneaked into the longitudinal mode
indeed must couple differently form the transversal parts even so this
looks not so obvious. Note, $m_H \gg M_W$ requires $\sqrt{\lambda} \gg g$
(since both masses, $M_W$ and $m_H$ are $\propto v$).

As in Ref.~\cite{Gastmans:2011ks} the role of the decoupling theorem
is confused with the equivalence theorem, I think. It is the latter which is
relevant to fix the ``free'' subtraction constant in the DR. A
\STi fixes it unambiguously. As mentioned earlier, while the decoupling
theorem holds for QED and QCD, where masses and couplings are
independent parameters. It does not hold for the weak sector of the SM
because of the mass coupling relations, which requires the coupling to
be proportional to the mass. All this is remembered in more detail in
the text preceding this update.

Acknowledgment:\\
I thank Oliver B\"ar for helpful discussions and for carefully reading the manuscript.

%
%
%
\end{document}